\documentstyle[12pt]{article}
\newcommand {\s}    {\sigma}

\newcommand {\jn}   {\uparrow}
\newcommand {\ur}   {\downarrow}

\def\normal{{\bigcirc \!\hskip -5pt n}} 

\def\0barra{{\rm O} \!\hskip -3.7pt {\rm l} } 

\def\1barra{1\! \hskip -1.1pt {\rm l}}

\title{ Free Fermions at Finite Temperature: An Application of 
    the Non-Commutative Algebra }

\vspace{-0.5cm}

\author{   S.M. de Souza$^{(1)}$, 
   O. Rojas Santos$^{(2)}$ \hspace{3pt} and \hspace{3pt} 
           M.T. Thomaz$^{(2)}$ \thanks{Corresponding author: 
Dr. Maria Teresa Thomaz;
 R. Domingos S\'avio Nogueira Saad n.$\!\!^{\rm o}$ 120  apto 404, 
 Niter\'oi, R.J., 24210--310, Brazil
 --Phone: (21) 620-6735; Fax: (21) 620-3881;  {\it E-mail}: mtt@if.uff.br.
}  \\
\\
\baselineskip =10pt
{ \small \it $^{(1)}$ Departamento de Ci\^encias Exatas \vspace{-0.2cm}}\\
{ \small \it Universidade Federal de Lavras \vspace{-0.2cm}}\\
{\small \it Caixa Postal  37 \vspace{-0.2cm} }\\
{ \small \it CEP: 37200-000, Lavras, MG,  Brazil }\\
{ \small \it $^{(2)}$ Instituto de F\'\i sica \vspace{-0.2cm}}\\
{ \small \it Universidade Federal Fluminense  \vspace{-0.2cm}}\\
{\small \it Av. Gal. Milton Tavares de Souza s/n.$\!\!^\circ$, 
	\vspace{-0.2cm} }\\
{ \small \it CEP: 24210-340, Niter\'oi, RJ, Brazil }   \\  }

\date{}

\begin{document}

\maketitle

\begin{abstract}

\baselineskip=14pt

 Charret {\it et. al.} applied the  properties of the 
Grassmann generators  to develop a
 new method to calculate the coefficientes of the high 
temperature expansion of the grand canonical partition function
 of   self-interacting fermionic models in any $d$-dimensions  \break
($d \geq 1$). The method explores the anti-commuting nature
 of fermionic fields and avoids the calculation 
of the fermionic path integral. 
We apply this new method to the relativistic free Dirac fermions
and recover the known results in the literature.

\end{abstract}

\vfill


\vspace{0.3cm}


\newpage

\baselineskip=18pt

%%%%%%%%%%%%%%%%%%%%%%%%%%%%%%%%%%%%%%%%%%%%%%%%%%%%%%%%%%%%%%%%%
%        INTRODUCAO
%%%%%%%%%%%%%%%%%%%%%%%%%%%%%%%%%%%%%%%%%%%%%%%%%%%%%%%%%%%%%%%%%

\section{ Introduction}

The path integral approach has been extensively applied to
calculate the thermodynamic properties of the quantum field
theories\cite{bernard,jackiw,Kapusta}. Through this approach,
it has been calculated the  leading contribution, 
in the high temperature limit ($\beta \ll 1$, where 
$\beta = \frac{1}{kT}$), to the effective potencial of these 
theories\cite{jackiw, yosef}. We also have a standard high
temperature perturbation theory derived from the path 
integral expression of the  partition 
function\cite{henderson,singh}. In the calculation of the 
Helmholtz free energy, via the path integral approach, we do not
keep track of its contribution that are $\beta$-independent.
In general this contribution is irrelevant to the 
thermodynamic properties of the model, but even we are 
not going to deal here with supersymmetric models, we remember
that this contributions are crucial to verify if the
supersymmtry is broken or not at finite 
temperature\cite{das,mishra,kumar}, since for a theory
to be supersymmetric the value of its vacuum energy
has to be zero.

For theories involving bosons, in the path integral we first
integrate over the conjugate momenta. Bernard\cite{bernard}
showed that in this case we have to be carefull in constructing
the path integral to get the overall $\beta$-dependent 
constant of the grand canonical partition function. This
constant does not appear in pure fermionic models since
in these cases the momentum conjugate to the fermionic
fields are their own hermitian conjugate. Even for these
fermionic models, along the calculation of the partition 
function done via the path integral  approach, 
we get  $\beta$-independent terms that are dropped out 
(see the free fermion case that is fully discussed in
reference \cite{Kapusta}).

The partition function of free fermions models is 
calculated directly from its  path  integral expression since it 
is equal to the determinant of its dynamical 
operator \cite{Kapusta, rothe}. When the 
lagrangean density of model has self-interacting 
fermionic terms, it is not possible any more to calculate
exactly the non-gaussian non-commutative path integral.
One common  approach is to get the bosonized version of the
model\cite{boson}. Another standard way to handle the 
path integral over non-commutative function 
is to do a perturbation theory taking care of the signs
 coming from fermion  loops\cite{bernard,jackiw,Kapusta}.

The interesting properties of the 
non-commutative Grassmann algebra
has been applied to get the contributions from 
spin configurations to the partition function of the 
classical bidimensional Ising model\cite{ising}.
In reference \cite{irazieJM}  Charret {\it et al.} proposed
a new way to calculate the coefficients of the high 
temperature expansion of the grand canonical partition 
function of self-interacting fermionic models in 
$d$-dimensions ($d \geq 1$). They applied the method
to the Hatsugay-Kohmoto model\cite{irazieJM}, that
is an exactly solvable model.  The approach was also
applied to the unidimensional Generalized Hubbard model
to get the coefficents up to order $\beta^3$ of the
high temperature expansion of its grand canonical 
partition function\cite{onofre}. Differently from other
approaches, these  coefficients calculated by this 
method are analytical and  exact.

Up to now the approach  of Charret {\it et al.} has only 
been applied to fermionic models already regularized 
 on a lattice with space unit one\cite{irazieJM,onofre}. 
The aim of the present paper is to use this method
to calculate the Helmholtz free energy of the free 
fermion Dirac, that is a continous theory  and 
whose exact result is already known in the 
literature\cite{Kapusta}. It is also important 
to check  its application to a fermionic lattice 
model\cite{rothe}.

In section 2 we give a summary of the results of the 
method of Charret {\it et al.} fully described in 
reference \cite{irazieJM}. In section 3 we apply 
the approach to the free Dirac fermion using two
expansions: in subsection 3.1 we first expand the
fermionic field operators in the basis of the eigenstates
of energy  and in subsection 3.2 we consider the 
naive fermion model on a three dimensional 
space lattice. In both ways, we show that is possible 
to re-sum the high temperature expansion of the
grand canonical partition function of the model and 
compare our results with the known ones in the
literature\cite{Kapusta, rothe}. In appendix
A we present the formulae related to the lagrangean
density of the free Dirac fermion. In appendix B
we have a dictionary of some formulae showing
their continuun and discrete expressions.

%%%%%%%%%%%%%%%%%%%%%%%%%%%%%%%%%%%%%%%%%%%%%%%%%%%%%%%%%%%%%%%%%
%        TERMINO DA INTRODUCAO
%%%%%%%%%%%%%%%%%%%%%%%%%%%%%%%%%%%%%%%%%%%%%%%%%%%%%%%%%%%%%%%%%

%%%%%%%%%%%%%%%%%%%%%%%%%%%%%%%%%%%%%%%%%%%%%%%%%%%%%%%%%%%%%%%%%
%        REVISAO DOS RESULTADOS CONHECIDOS
%%%%%%%%%%%%%%%%%%%%%%%%%%%%%%%%%%%%%%%%%%%%%%%%%%%%%%%%%%%%%%%%%

\section{  A Survey of the  Approach of Charret {\it et al.}}

The grand canonical partition function of any quantum system can be 
written as a trace  over all quantum physical 
states\cite{Kapusta,reif}

\begin{equation}
{\cal Z}(\beta)=Tr[e^{-\beta{\bf K}}],  \label{1}
\end{equation}

\noindent where $\beta=\frac{1}{kT}$, $k$ is the Boltzmann's constant
and $T$ is the absolute temperature. The operator ${\bf K}$ is  defined
as ${\bf K}={\bf H}-\mu{\bf N}$, ${\bf H}$ being the   hamiltonian
of the system, $\mu$ is the chemical potential and
 ${\bf N}$ is a conserved operator. In the high temperature
 limit $(\beta\ll 1)$, ${\cal Z}(\beta)$ has the
expansion

\begin{equation}
{\cal Z}(\beta)=Tr[\1barra]+\sum_{n=1}^{\infty}
\frac{(-\beta)^n}{n!}Tr[{\bf K}^n].  
        \label{2}
\end{equation}

\noindent The spectation value  of any fermionic operator
 can be written as  a multivariable integral over Grassmann 
variables\cite{itzykson,irazieA}. The mapping among
the fermionic operators that appear in the fermionic 
model  and the grassmannian generators is 
such that they satisfy the same algebra.  Let ${\bf a}_i^\dagger$
 and  ${\bf a}_j$  be the hermitian conjugate fermionic  operators
 that satisfy the anti-commuting relations

\begin{equation}
\{ {\bf a}_i, {\bf a}^\dagger_j  \} = \1barra \, \delta_{ij}
\hspace{1cm} {\rm and} \hspace{1cm} 
\{ {\bf a}_i, {\bf a}_j \} =0, \label{3}
\end{equation}

\noindent where $i= 1,..,{\cal N}$.
The generators of the associated  Grassmann algebra has dimension 
$2^{2{\cal N}}$, and can be  written explicitly as 
$\{ \bar{\eta}_1, \cdots , 
\bar{\eta}_{\cal N}; \eta_1, \cdots, \eta_{\cal N} \}$.  
They satisfy the following anti--commutation relations:  

\begin {equation}
\{{\eta}_i , \eta_j \} = 0, \hspace {0.5cm} \{ \bar{\eta}_i ,\bar\eta_j
\} =0 \hspace {0.5cm} {\rm and} \hspace {0.5cm} \{ \bar{\eta}_i , \eta_j
\}=0.  \label{4}
\end{equation} 

\noindent The mapping 

\begin{equation}
 {\bf a}_i^\dagger \rightarrow \bar{\eta}_i 
\hspace{1cm} {\rm and}
\hspace{1cm} {\bf a}_j \rightarrow 
\frac{\partial}{ \partial\bar{\eta}_j}, \label{5}
\end{equation}

\noindent  preserves the algebra (\ref{3}) due to the 
algebra (\ref{4}) satisfied by the grassmannian generators.

For any self-interacting fermionic model in a $d$-dimension lattice 
($d \geq 1$), the coefficients of high temperature expansion 
(\ref{2}) can be written as the multivariable Grassmann
 integral\cite{irazieJM}

\begin{eqnarray}   
Tr[{\bf K}^n] & = &\int\prod_{I=1}^{2nN^d}d\eta_Id\bar\eta_I\;
e^{\sum_{I,J=1}^{2nN^d}\bar\eta_IA_{I,J}\eta_J}\times
\hskip7cm   \nonumber\\\
%
%segunda linha
%
& & \hspace{1cm}
\times{\cal K}^{\normal} (\bar\eta,\eta,\nu=0)
{\cal K}^{\normal} (\bar\eta,\eta,\nu=1)\dots
{\cal K}^{\normal} (\bar\eta,\eta,\nu=n-1),
    \label{6}
\end{eqnarray}

\noindent  $N^d$ is the number of sites in the $d$-dimensional 
lattice.  Matrix ${\bf A}$ is independent of the operators
${\bf H}$ and  ${\bf N}$ and is equal to

\begin{equation}
{\bf A}=\left( \begin{array} {c c}
{\bf A}^{\jn\jn} & \0barra \\
\0barra & {\bf A}^{\ur\ur}
\end{array}\right).    \label{7}
\end{equation}

\noindent Each element of matrix ${\bf A}$ is a matrix of
dimension ${nN^d\times nN^d}$ and $ \0barra$ is the null matrix
in this dimension. The indices  $I,J$ in matrix 
${\bf A}$ vary in the interval, $I,J=1,2,\dots,2nN^d$.  
The matrices  ${\bf A}^{\jn\jn}$ and  ${\bf A}^{\ur\ur}$ 
are  equal and

\begin{equation}   
{\bf A}^{\jn\jn}={\bf A}^{\ur\ur}=\left( \begin{array} {c c c c c}
\1barra_{N^d\times N^d} &  -\1barra_{N^d\times N^d} & 
\0barra_{N^d\times N^d} & \dots & \0barra_{N^d\times N^d}\\
\0barra_{N^d\times N^d} &  \1barra_{N^d\times N^d} & 
-\1barra_{N^d\times N^d} & \dots & \0barra_{N^d\times N^d}\\
\vdots & & & & \vdots\\
\1barra_{N^d\times N^d} & \0barra_{N^d\times N^d} &
 \0barra_{N^d\times N^d} & \dots & \1barra_{N^d\times N^d}
\end{array}\right) ,
    \label{8}
\end{equation}

\noindent where $\1barra_{N^d\times N^d}$ 
and  $\0barra_{N^d\times N^d}$ are the identity 
and null matrices of dimension ${N^d\times N^d}$
respectively. Due to the fact that the submatrices ${\bf A}^{\jn\ur}$ 
and ${\bf A}^{\ur\jn}$ are null, the result of the
 multivariable integral (\ref{6}) is equal to the product of the 
independent contributions of the sectors $\sigma\sigma=\jn\jn$ and
 $\sigma\sigma=\ur\ur$.

The grassmannian function ${\cal K}^{\normal}(\bar\eta,\eta)$ 
in eq.(\ref{6}) is obtained from the normal ordered operator
 ${\bf K}$ by doing the naive mapping:
${\bf a}_{i, \sigma}^{\dag}\rightarrow\bar\eta_I$ and
$ {\bf a}_{i, \sigma}\rightarrow\eta_I$, where $i=1,2,...,N^d$
 and $ \sigma = \uparrow, \downarrow$.

Finally, the result of the integrals that contribute to 
$Tr[{\bf K}^n]$ are independent of the $\sigma\sigma$
 sector since ${\bf A}^{\uparrow\uparrow} = 
{\bf A}^{\downarrow\downarrow}$.  We present here
only the results of the
integrals of the sector $\s\s=\jn\jn$. All those
integrals are of the following type: 

\begin{equation}\label{M(L,K)}
M(L,K)=\int\prod_{i=1}^{nN^d}d\eta_id\bar\eta_i\; \bar\eta_{l_1}\eta_{k_1}
\dots \bar\eta_{l_m}\eta_{k_m}\;
e^{\sum_{i,j=1}^{nN^d}\bar\eta_iA_{ij}^{\jn\jn}\eta_j},
    \label{9}
\end{equation}

\noindent where  $L\equiv\{l_1,\dots,l_m\}$ and $K\equiv\{k_1,\dots,k_m\}$.
We should remember that the grassmannians functions 
${\cal K}^{\normal}$ are polynomials of the generators of the algebra.
The products  $\eta\bar\eta$ are ordered such that 
$l_1<l_2<...<l_m$ and  $k_1<k_2<...<k_m$. The results of the 
integrals of type (\ref{9}) are\cite{irazieB}

\begin{equation} 
M(L,K)=(-1)^{\sum_{i=1}^{m}(l_i+k_i)}A(L,K),
   \label{10}
\end{equation}

\noindent where $A(L,K)$ is the determinant obtained from matrix
${\bf A}^{\jn\jn}$ by the cut of the lines $l_1,l_2,...,l_m$
and the columns $k_1,k_2,...,k_m$.  $M(L,K)$ is a co-factor
of matrix ${\bf A}^{\jn\jn}$. Result (\ref{10}) reduce the calculation
of the multivariable  integrals over anti-commuting variables 
to the calculation of co-factors of a well defined matrix
whose elements are commuting numbers.

\vspace{0.3cm}

In general it should not be easy to calculate the co-factors of
matrix ${\bf A}^{\s\s}$ since each of its elements is a matrix
of dimension  $N^d\times N^d$. However, due to the block
structure of matrix ${\bf A}^{\s\s}$  Charret {\it et. al}  
showed in reference \cite{irazieJM} that is possible
diagonalize it for any value of $n$ and $N$. The results are
analytical and allow us to take the thermodynamic limit.

%%%%%%%%%%%%%%%%%%%%%%%%%%%%%%%%%%%%%%%%%%%%%%%%%%%%%%%%%%%%%%%%%
%        TERMINO DA REVISAO DOS RESULTADOS CONHECIDOS
%%%%%%%%%%%%%%%%%%%%%%%%%%%%%%%%%%%%%%%%%%%%%%%%%%%%%%%%%%%%%%%%%

%%%%%%%%%%%%%%%%%%%%%%%%%%%%%%%%%%%%%%%%%%%%%%%%%%%%%%%%%%%%%%%%%
%        FUNCAO DE PARTICAO GRAN CANONICA PARA FERMIONS LIVRES
%%%%%%%%%%%%%%%%%%%%%%%%%%%%%%%%%%%%%%%%%%%%%%%%%%%%%%%%%%%%%%%%%

\section{Grand Canonical Partition Function for Free \hfill \break
Relativistic Fermions}

The lagrangean density of the free Dirac fermion is:

\begin{equation} 
{\cal L}=\bar\Psi(\vec x,t)(\partial \hspace{-0.25cm}/ 
 \hspace{0.17cm}  - m)\Psi(\vec x,t),
       \label{11}
\end{equation}

\noindent whose fermionic field operators satisfy the anti-commutation
ralations

\vspace{-0.5cm}

\begin{equation}
\{\Psi_{\mu}(\vec x,t),\Pi_{\nu}(\vec x^{\prime},t)\}  = 
i\delta_{\mu\nu}\delta(\vec x-\vec x^{\prime}),
\hspace{1cm} \mu,\nu=1,2,3,4   \label{12}  
\end{equation}

\noindent and 

\vspace{-0.6cm}

\begin{equation}
\{\Psi_{\mu}(\vec x,t),\Psi_{\nu}(\vec x^{\prime},t)\}  = 
\{\Pi_{\mu}(\vec x,t),\Pi_{\nu}(\vec x^{\prime},t)\}=0,
      \label{13}
\end{equation}

\vspace{0.3cm}

\noindent where $\Pi_{\nu}(\vec x,t)$ is the canonical 
momentum of $\Psi_{\nu}(\vec x,t)$ and 
$\Pi_{\nu}(\vec x,t)=i\Psi_{\nu}^{\dag}(\vec x,t)$.

From the  lagrangean density (\ref{11}) and the equation of 
motion satisfied by the fermionic  field operators,
the total hamiltonian  operator of the system can be written as 

\begin{equation}
{\bf H}=\int_{V}d^3{\vec x}\;i\Psi^{\dag}(\vec x,t)\partial_0\Psi(\vec x,t).
    \label{14}
\end{equation}

We use the natural units where $\hbar=c=e=1$. Our metric is 
 $diag(g_{\mu\nu})=(1,-1,-1,-1)$.

%%%%%%%%%%%%%%%%%%%%%%%%%%%%%%%%%%%%%%%%%%%%%%%%%%%%%%%%%%%%%
% Expansao na base dos autoestados de energia
%%%%%%%%%%%%%%%%%%%%%%%%%%%%%%%%%%%%%%%%%%%%%%%%%%%%%%%%%%%%%

\subsection{Expansion in the Basis of Eigenstates of Energy}

The fermionic operator $\Psi(\vec x,t)$ written in the basis
 of eigenstates of energy of the free Dirac fermion is

\begin{equation}
\Psi(\vec x,t)=\frac{1}{V^{1/2}}\sum_{\vec k}
\sum_{r=1}^{2}[{\bf a}_r(\vec k)u_r(\vec k)
e^{-ik_{\mu}x^{\mu}}+{\bf b}_r^{\dag}(\vec k)v_r(\vec k)
e^{ik_{\mu}x^{\mu}}],
    \label{15}
\end{equation}

\noindent where $k_{\mu}x^{\mu}=k_0 x_0-\vec k.\vec x$, 
and  $k_0=\sqrt{|\vec k|^2+m^2}>0$. 
The destruction fermionic operators  ${\bf a}_r$, ${\bf b}_r$ 
and their respective hermitian conjugate satisfy the anti-commutation
relations

\begin{equation}
\{{\bf a}_r(\vec k),{\bf a}_s^{\dag}(\vec k^{\prime})\}=
\delta_{rs}\delta_{\vec k,\vec k^{\prime}}
\hspace{1cm}  {\rm and} \hspace{1cm}
\{{\bf b}_r(\vec k),{\bf b}_s^{\dag}(\vec k^{\prime})\}=\
\delta_{rs}\delta_{\vec k,\vec k^{\prime}}.
       \label{16}
\end{equation}

\noindent All others anti-commutation relations of these 
operators are null. The spinors components $u_r(\vec k)$
 and $v_r(\vec k)$ are given by eqs.(\ref{A4}) and
(\ref{A5}).

The hamiltonian operator written in terms of creation and 
destruction operators of defined energy becomes

\begin{equation}
{\bf H}=- \1barra\sum_{\vec k}\sum_{r=1}^{2}k_0+\sum_{\vec k}
\sum_{r=1}^{2}k_0[{\bf a}_r^{\dag}(\vec k){\bf a}_r(\vec k)+
{\bf b}_r^{\dag}(\vec k){\bf b}_r(\vec k)],
     \label{17}
\end{equation}

\noindent $\1barra$ been the identity operator. We define 
$E_0$ as the vacuum energy,
$E_0\equiv\langle 0|{\bf H}|0\rangle=-2\sum_{\vec k}k_0$, 
 $|0\rangle$ being the vacuum state of the fermionic model.

\vspace{0.2cm}

Our aim is to calculate the grand canonical partition function
of the free Dirac fermion in contact with a reservoir of heat and
electric charge. Eqs. (\ref{6}), (\ref{9}) and (\ref{10})
 allow us to calculate the coefficients of the high temperature
expansion of this function. For the present case, the operator
${\bf K}$ in expression (\ref{1}) is: 
 ${\bf K}=:{\bf H}:+E_0 \1barra -\mu{\bf Q}$, where $\mu$
is the chemical potential and  ${\bf Q}$ is the total electric
charge operator of the free relativistic fermions. Operator
 ${\bf Q}$ written in terms of the creation and destruction 
operators is 

\vspace{-0.5cm}

\begin{eqnarray}
{\bf Q} & = & 2\sum_{\vec k}  \1barra +
 \sum_{\vec k}\sum_{r=1}^{2}
[{\bf a}_r^{\dag}(\vec k){\bf a}_r(\vec k)-
{\bf b}_r^{\dag}(\vec k){\bf b}_r(\vec k)].  \label{19}   \\
%
%segunda linha
%
 & \equiv &  Q_0 \1barra + :{\bf Q}:  \; .\nonumber
\end{eqnarray}

Finally, we have that ${\bf K}$ operator can be written
as

\vspace{-0.5cm}

\begin{eqnarray}
{\bf K} & = &\sum_{\vec k}\sum_{r=1}^{2}(k_0-\mu){\bf a}_r^{\dag}(\vec k)
{\bf a}_r(\vec k)+\sum_{\vec k}\sum_{r=1}^{2}(k_0+\mu){\bf b}_r^{\dag}
(\vec k){\bf b}_r(\vec k)+ \1barra (E_0-\mu Q_0)\nonumber\\
%
%segunda linha
%
& \equiv & {\bf K}_a+{\bf K}_b+ \1barra (E_0-\mu Q_0).
     \label{20}
\end{eqnarray}

\noindent Due to the anti-commutation relations (\ref{16}), we 
have that $[{\bf K}_a,{\bf K}_b]=0$, and therefore
 ${\cal Z}(\beta)$ can be  written as

\vspace{-0.3cm}

\begin{equation}
{\cal Z}(\beta)= e^{-\beta(E_0-\mu Q_0)}{\cal Z}_a(\beta).
{\cal Z}_b(\beta), 
     \label{21}
\end{equation}

\noindent where

\vspace{-0.5cm}

\begin{equation}
{\cal Z}_a(\beta)\equiv Tr_a[e^{-\beta{\bf K}_a}]
\hspace{1cm}  {\rm and} \hspace{1cm}
{\cal Z}_b(\beta)\equiv Tr_b[e^{-\beta{\bf K}_b}].
     \label{22}
\end{equation}

\noindent The calculation of the functions
  ${\cal Z}_a(\beta)$ and   ${\cal Z}_b(\beta)$ 
are equivalent. We  present here only
 the details  of the calculation of ${\cal Z}_a(\beta)$.
 For a free Dirac fermion, the traces for 
$r =1$  and  $r = 2$  are equal.  Then, 

\begin{eqnarray}
{\cal Z}_a(\beta) & = &  Tr_a[e^{-\beta{\bf K}_a}]=[Tr_{a}^{(1)}
[e^{-\beta{\bf K}_{a}^{(1)}}]]^2  \nonumber \\
%
%segunda linha
%
 &  =  &  
\left[ \prod_{\vec k}
Tr_{a}^{(1)}[e^{-\beta(k_0-\mu){\bf n}_{a}^{(1)} (\vec k)}] \right]^2.
     \label{23}
\end{eqnarray}

\noindent In $Tr_{a}^{(1)}$ we have the vector 
$\vec{k}$ fixed. Since ${\bf n}_{a}^{(1)} (\vec k)$ 
is a commuting operator, we apply the Newton 
multinomial to write 

\begin{equation} 
Tr_{a}^{(1)} [e^{-\beta(k_0-\mu){\bf n}_{a}^{(1)} (\vec k)}]=
Tr_{a}^{(1)} [\1barra]+\sum_{n=1}^{\infty}\frac{(-\beta)^n}{n!}(k_0-\mu)^n
Tr_{a}^{(1)}[{\bf n}_{a}^{(1)} (\vec k)^n].
      \label{24}
\end{equation}

\noindent We should note that for fixed $\vec{k}$
the anti-commutation relations (\ref{16}) are
 identitical to the relations (\ref{3}), therefore, 
$Tr_{a}^{(1)} [{\bf n}_{a}^{(1)} (\vec k)^n]$
can be written as a Grassmann multivariable 
integral (\ref{6}) with an equivalent mapping  to eq.(\ref{5})
for the associated generators of the non-commutative
 algebra. The traces that contribute to eq.(\ref{24}) 
are written as the following anti-commuting integrals:

\begin{equation} 
Tr_{a}^{(1)} [ \1barra] = \int\prod_{I=1}^{n} d\eta_{\vec k}(I)
 d\bar\eta_{\vec k}(I) 
e^{\sum_{I,J=1}^{n}\bar\eta_{\vec k}(I)A_{IJ}^{(11)}
\eta_{\vec k}(J)},    \label{25} 
\end{equation}

\noindent and
\vspace{-0.5cm}

\begin{eqnarray}
Tr_{a}^{(1)} [{\bf n}_{a}^{(1)} (\vec k)^n] & = &\int\prod_{I=1}^{n} 
d\eta_{\vec k}(I) d\bar\eta_{\vec k}(I) e^{\sum_{I,J=1}^{n}
\bar\eta_{\vec k}(I)A_{IJ}^{(11)}\eta_{\vec k}(J)}\times\nonumber\\
%
%segunda linha
%
&  & \hspace{-1cm}
 \times \bar\eta_{\vec k}(1)\eta_{\vec k}(1)\dots\bar
\eta_{\vec k}(n)\eta_{\vec k}(n), \hspace{1.5cm} {\rm for} 
\;\; (n>0). 
    \label{26}
\end{eqnarray}

\noindent The grassmannian expression for
$Tr_{a}^{(1)} [{\bf n}_{a}^{(1)} (\vec k)^n]$ is obtained from 
eq.(\ref{6}) with  $N = 1$ and $d=1$.  Therefore the elements
of matrix ${\bf A}^{(11)}$ in eq.(\ref{8})
are just numbers. From eqs.(\ref{9}) and (\ref{10}),  
we conclude that having the $\bar{\eta}$'s
in the integrand corresponds to deleting the first $n$
{\em lines} in matrix ${\bf A}^{(11)}$ and, in the same way,
having the $\eta$'s corresponds to deleting the first $n$
{\em columns} of the same matrix.  Therefore, from the expression
of matrix ${\bf A}^{(11)}$ (see eq.(\ref{8})), for
arbitrary $n$, we realize that the matrix
${\bf A}^{(11)}_{n}$ obtained after the cuts of 
 $n$-first lines and $n$-first collumns is an upper
triangular matrix whose determinant is equal to 1. 
Beside this, we have that 
$ \det [ {\bf A}^{(11)}] =2$, for any value of $n$.
In summary, for arbitrary $n$ we have that

\begin{equation}
Tr_{a}^{(1)} [ \1barra] =2  
\hspace{1cm} {\rm and} \hspace{1cm}
Tr_{a}^{(1)} [{\bf n}_{a}^{(1)} (\vec k)^n] =1.
   \label{27}
\end{equation}

\noindent Substituting results (\ref{27}) in expression (\ref{26})
and resumming it we get

\begin{equation}
Tr_{a}^{(1)} [e^{-\beta{\bf K}_{a}^{(1)}}] =
\prod_{\vec k}(1+e^{-\beta(k_0-\mu)}),
    \label{28}
\end{equation}

\noindent that returning to eq.(\ref{21}) gives us

\begin{equation}
{\cal Z}(\beta)=[\prod_{\vec k}e^{2 \beta(k_0+\mu)}
(1+e^{-\beta(k_0-\mu)})(1+e^{-\beta(k_0+\mu)})]^2.
   \label{29}
\end{equation}

The relation between the  Helmholtz free energy and the grand 
canonical partition function is\cite{reif}

\begin{equation}
{\cal W}(\beta)=-\frac{1}{\beta}\ln({\cal Z}(\beta)).
    \label{30}
\end{equation}

\noindent Substituting result  (\ref{29}) in eq.(\ref{30}) 
the expression derived for the Helmholtz free energy is 

\begin{equation}\label{potencial}
{\cal W}(\beta)=-\frac{2}{\beta}\sum_{\vec k}[\beta(k_0+\mu)+
\ln(1+e^{-\beta(k_0-\mu)})+\ln(1+e^{-\beta(k_0+\mu)})].
    \label{31}
\end{equation}

%%%%%%%%%%%%%%%%%%%%%%%%%%%%%%%%%%%%%%%%%%%%%%%%%%%%%%%%%%%%%
% Termino da Expansao na base dos autoestados de energia
%%%%%%%%%%%%%%%%%%%%%%%%%%%%%%%%%%%%%%%%%%%%%%%%%%%%%%%%%%%%%

%%%%%%%%%%%%%%%%%%%%%%%%%%%%%%%%%%%%%%%%%%%%%%%%%%%%%%%%%%%%%
% Fermions de Dirac livres na rede 
%%%%%%%%%%%%%%%%%%%%%%%%%%%%%%%%%%%%%%%%%%%%%%%%%%%%%%%%%%%%%

\subsection{Free Dirac Fermions on the Lattice}

The lattice calculation of models including fermions has been an 
important tool to learn the properties of these models. The
aim of this subsection  is to show that the method of 
Charret {\it et al.} can be equally well applied to the lattice
version of fermionic models. To do so, we consider the most
naive lattice realization of the free Dirac fermion\cite{rothe}.
We remember that the crucial point to apply the method of 
Charret {\it et al.} is to work with fermionic field
operators that satisfy  the anti-commutation
relations  (\ref{3}).

The hamiltonian operator of the free Dirac fermion is

\vspace{-0.3cm}

\begin{equation}
{\bf H}=\int_{V}d^3{\vec{x}}\;\;  i\Psi^{\dag}(\vec{x},t)
[ \gamma^0 \vec{\gamma} \cdot \vec{\nabla} - m\gamma^0]
\Psi(\vec x,t),
    \label{32}
\end{equation}

\noindent where $\vec{\gamma} = (\gamma^1, \gamma^2, \gamma^3)$.
From appendix B, we get that the  operator ${\bf H}$ 
written on the lattice  becomes 

\vspace{-0.5cm}

\begin{eqnarray}
{\bf H} & = &\sum_{n_1, n_2, n_3 = -\frac{N}{2}}^{\frac{N}{2}} 
\hspace{4pt} \sum_{\alpha, \beta = 1}^{4} \hspace{2pt}  \sum_{j=1}^{3}
\left\{  i\frac{a^2}{2} \Psi_\alpha^{\dag} (\vec{n}a,t)
 (\gamma^0 \gamma^j)_{\alpha \beta}
\left[ \Psi_\beta (\vec{n}a + \hat{\jmath}a,t) 
- \Psi_\beta (\vec{n}a - \hat{\jmath}a,t)  \right] - \right.   
                         \nonumber \\
%
%segunda linha
%
 & &    \hspace{3cm}
 - \left. ia^3 m \Psi_\alpha^{\dag} (\vec{n}a,t)  
  (\gamma^0)_{\alpha \beta} \Psi_\beta (\vec{n}a,t) \right\},
    \label{33}
\end{eqnarray}

\noindent where $a$ is the distance between the nearest sites
in each space direction. The space point $\vec{x}$ on the lattice 
is written as: $\vec{x}= \vec{n} a$. $N$ is the total
 number of space sites in each direction. 
For simplicity we take to be $N$ even.

Operator ${\bf K}$ in expression (\ref{1}) is: ${\bf K} =
{\bf H} - \mu{\bf Q}$, where $\mu$ is the chemical potential
and ${\bf Q}$ is the total electric charge operator. The 
discrete expression of ${\bf Q}$ is

\begin{equation}
 {\bf Q} = a^3 \hspace{-0.5cm}
 \sum_{n_1, n_2, n_3 = -\frac{N}{2}}^{\frac{N}{2}}
\hspace{4pt} \Psi_\alpha^{\dag} (\vec{n}a,t)
\Psi_\alpha (\vec{n}a,t).
    \label{34}
\end{equation}

\vspace{0.2cm}

Imposing the space periodic boundary condition for
the fermionic field operators, their Fourier decompositions are

\begin{equation}
\Psi_\alpha (\vec{n}a,t) = \frac{1}{\sqrt{V}} 
\sum_{k_1, k_2, k_3 = -\frac{N}{2}}^{\frac{N}{2}-1}
\tilde{\psi}_\alpha (\frac{\pi \vec{k}}{L},t)
e^{i \frac{\pi}{L}\vec{k}\cdot\vec{n}a },
    \label{35}
\end{equation}

\noindent where $\alpha= 1,2,3,4$ and  $L = \frac{Na}{2}$. The 
fermionic field operators $\Psi_\alpha (\vec{x},t)$
and $\Psi_\alpha^{\dag} (\vec{x},t)$ satisfy the
anti-commutation relations (\ref{12}) and (\ref{13})
which imply  that the Fourier components of the 
fermionic field operators obey the relations:

\begin{equation}
\{ \tilde{\psi}_\alpha (\frac{\pi \vec{l}}{L},t), 
\tilde{\psi}_\beta^{\dag} (\frac{\pi \vec{k}}{L},t) \} 
=  \delta_{\alpha \beta} \; \delta_{_{\vec{l}, \vec{k}}}^{^{(3)}}
\hspace{0.5cm} {\rm and} \hspace{0.5cm}
\{ \tilde{\psi}_\alpha (\frac{\pi \vec{l}}{L},t), 
\tilde{\psi}_\beta (\frac{\pi \vec{k}}{L},t) \} = 0, 
   \label{36}
\end{equation}

\noindent that are identical to relations (\ref{3}), the
necessary algebra to apply the method of Charret {\it et al.}.

\vspace{0.2cm}

In momentum space, operator ${\bf K}$ is written as

\begin{equation}
{\bf K} \equiv  \frac{1}{a} \;  
\sum_{l_1, l_2, l_3 = - \frac{N}{2}}^{ \frac{N}{2} - 1 }
 {\bf K}_{\vec{l}} \,,
    \label{37}
\end{equation}

\noindent where 

\vspace{-0.5cm}

\begin{equation}
{\bf K}_{\vec{l}} \equiv  
\tilde{\Psi}^{\dag} \left( \frac{\pi}{L} \vec{l}, t\right) \;
{\bf R}  \left( \frac{\pi}{L} \vec{l}\right) \;
\tilde{\Psi} \left( \frac{\pi}{L} \vec{l}, t\right) ,
     \label{38}
\end{equation}

\noindent and 

\begin{equation}
\tilde{\Psi} \left( \frac{\pi}{L} \vec{l}, t\right) \equiv
\pmatrix{ \tilde{\psi}_1  \cr
          \tilde{\psi}_2   \cr
          \tilde{\psi}_3   \cr
          \tilde{\psi}_4   }.
     \label{39} 
\end{equation}

\noindent The matrix ${\bf R}  \left( \frac{\pi}{L} \vec{l}\right)$
is defined as

\begin{equation}
{\bf R}  \left( \frac{\pi}{L} \vec{l}\right) = 
\pmatrix{ a ( m - \mu) \1barra & 
       sin\left(\frac{\pi}{L} l_j a \right) \sigma_j   \cr
          sin\left(\frac{\pi}{L} l_j a \right) \sigma_j  &
       -a ( m + \mu) \1barra   }.
    \label{40}
\end{equation}

\noindent We have an implicit sum over $j$ in the
off-diagonal  elements of  matrix ${\bf R}$ and
 $\1barra$ is the identity matrix of 
dimension $2\times 2$. Due to the anti-commutation relations
(\ref{36}) we have that 
$[{\bf K}_{\vec{l}}, {\bf K}_{\vec{k}} ] =0$. Therefore, the 
grand canonical partition function of the model becomes

\begin{equation}
{\cal Z}(\beta) = \prod_{l_1, l_2, l_3= -\frac{N}{2}}^{\frac{N}{2}-1}
\;\,  Tr_{\vec{l}} \; 
\left[ e^{-\frac{\beta}{a} \;{\bf K}_{\vec{l} } } \right] ,
    \label{41}
\end{equation}

\noindent where $ Tr_{\vec{l}}$ means that the trace is calculated for 
fixed $\vec{l}$. We should notice that the operator 
${\bf K}_{\vec{l} }$ is not diagonal in momentum space. We make
the similarity transformation:
 ${\bf P} {\bf R} {\bf P}^{-1} = {\bf D}$, where the diagonal 
matrix ${\bf D}$ is 

\begin{equation}
{\bf D} = \pmatrix{  \lambda_+ \1barra  &  \0barra  \cr
                     \0barra  &  \lambda_- \1barra  } ,
    \label{42}
\end{equation}

\noindent where $\lambda_+$ and $\lambda_-$ are the eigenvalues
of matrix ${\bf R}$, and

\begin{equation}
\lambda_\pm =- a (\mu \pm  \tilde{\omega} (\vec{l}) )
       \label{43}
\end{equation}

\noindent with

\begin{equation}
 \tilde{\omega} (\vec{l}\,) \equiv
\sqrt{  m^2 + \tilde{p_1}^2 + \tilde{p_2}^2 + \tilde{p_3}^2},
     \label{43.1}
\end{equation}

\noindent and $\vec{l} \equiv ( l_1, l_2, l_3)$. 
We followed reference \cite{rothe} and made the
 change of variables

\begin{equation}
 \tilde{p}_i = \frac{1}{a} sin(p_i a),
      \label{46}
\end{equation}

\noindent where $p_i= \frac{\pi}{L} l_i$. The new fermionic 
field operators $\Psi^\prime = {\bf P} \tilde{\Psi}$ 
and ${\Psi^\prime}^{\dag} =  \tilde{\Psi}^{\dag} {\bf P}^{-1}$
also preserves the anti-commutation relations (\ref{36}). 

The function ${\cal Z}(\beta)$ written in terms of the new 
fermionic field operators has the same form as the r.h.s. of 
eq.(\ref{23}). Following similar steps, we get  the
Helmholtz free energy for the model on the lattice, that is, 

\begin{eqnarray}
{\cal W} (\beta) & = & - 2 \sum_{l_1, l_2, l_3 = 
-\frac{N}{2}}^{\frac{N}{2} - 1}
\left( \mu + \tilde{\omega} (\vec{l}\,)  \right)- 
                  \nonumber  \\
%
% segunda linha
%
&  & \hspace{-1cm}
-\frac{2}{\beta} 
\sum_{l_1, l_2, l_3 = -\frac{N}{2}}^{\frac{N}{2} - 1}
\left[ ln\left( 1 + 
e^{-\beta( \tilde{\omega} (\vec{l}\,) + \mu )} \right)+
ln\left( 1 + 
e^{-\beta( \tilde{\omega} (\vec{l}\,) - \mu )} \right)
    \right] .   \label{44}
\end{eqnarray}

%   \label{45}

\noindent In the limit $a \rightarrow 0$ the function 
${\cal W} (\beta)$ agrees with the result of reference
\cite{rothe} and is equal to twice the result (\ref{31}).
Here as well  in the usual lattice calculation with 
fermions, the doubling problem is lifted by including 
the  Wilson term

\begin{equation}
{\bf H}^{(W)} = i \frac{ra}{2} \int_V\; d^3 \vec{x} \;
\Psi^{\dag} (\vec{x},t) \, \gamma^0 \,\nabla^2 \Psi(\vec{x},t),
   \label{47}
\end{equation}

\noindent in hamiltonian (\ref{32})  and $r$ is the
 Wilson's constant.

%%%%%%%%%%%%%%%%%%%%%%%%%%%%%%%%%%%%%%%%%%%%%%%%%%%%%%%%%%%%%
% Termino de Fermions de Dirac livres na rede 
%%%%%%%%%%%%%%%%%%%%%%%%%%%%%%%%%%%%%%%%%%%%%%%%%%%%%%%%%%%%%

%%%%%%%%%%%%%%%%%%%%%%%%%%%%%%%%%%%%%%%%%%%%%%%%%%%%%%%%%%%%%%%%%%%%%
%    TERMINO DA FUNCAO DE PARTICAO GRAN CANONICA PARA FERMIONS LIVRES
%%%%%%%%%%%%%%%%%%%%%%%%%%%%%%%%%%%%%%%%%%%%%%%%%%%%%%%%%%%%%%%%%%%%%

%%%%%%%%%%%%%%%%%%%%%%%%%%%%%%%%%%%%%%%%%%%%%%%%%%%%%%%%%%%%%%%%%
%      CONCLUSOES
%%%%%%%%%%%%%%%%%%%%%%%%%%%%%%%%%%%%%%%%%%%%%%%%%%%%%%%%%%%%%%%%%

\section{Conclusions}

Recently Charret {\it et al.} proposed   a new way to calculate
the coefficients of the high temperature expansion of the 
grand canonical partition function ${\cal Z}(\beta)$
of any self-interacting fermionic model in $d$-dimension
($d\geq 1$)\cite{irazieJM}. To apply this method 
to calculate these coefficients, it is
enough to write the second quantization expression of the 
hamiltonian  and a conserved operator in terms of operators
that satisfy the anti-commutation relations (\ref{3}). 
In this approach, at each order $\beta^n$ of the high temperature 
expansion of  the function ${\cal Z}(\beta)$, the calculation of
the coefficients is reduced to get the co-factors
of a matrix with commuting entries (see matrix ${\bf A}$ in
eqs. (\ref{7}) and (\ref{8})). For a fixed number of sites on the 
lattice, all the mathematical objects in the calculation are 
well-defined. This approach avoids to calculate the fermionic 
path integral of ${\cal Z}(\beta)$.

We applied the method of Charret {\it et al.} to the free 
Dirac fermion by first expanding it in the basis of the 
eigenstates of energy of the free fermions. 
The hamiltonian  and  the  total electric 
charge operators (eqs.(\ref{14}) and (\ref{19}) 
respectively) include the respective vacuum contribution.
Result (\ref{31}) gives two divergent terms to the
Helmholtz free energy ${\cal W}(\beta)$: the vacuum energy
 and the electric charge of the vacuum. This last divergent term does 
not appear when we do the calculation via the
usual path integral approach\cite{Kapusta}, while the 
$\beta$-dependent terms in ${\cal W}(\beta)$ are identical
in both methods. In the path integral calculation\cite{Kapusta}
divergent terms that are $\beta$ and $\mu$-independent are dropped 
out. This does not happens in the present approach. 

Since the calculation of fermionic lattice models are an 
important tool, we also applied the approach of
 Charret {\it et al.} to a naive version of the
free Dirac fermion on the lattice\cite{rothe}. In this case we 
also get the doubling problem that is lifted  by the Wilson term,
as happens in the usual fermion lattice calculations. 

In summary, we can affirm that the method of Charret{\it et al.}
can also be applied to continuous fermionic models. It is an
analytical approach that allows without ambiguity 
to calculate all terms
of the function ${\cal Z}(\beta)$ for the free Dirac fermion, 
including the divergent terms coming from the vacuum contribution.
The next step is to study  in this approach the renormalization 
scheme associated to physical quantities for fermionic models
with self-interacting terms.

%%%%%%%%%%%%%%%%%%%%%%%%%%%%%%%%%%%%%%%%%%%%%%%%%%%%%%%%%%%%%%%%%
%      TERMINO DAS CONCLUSOES
%%%%%%%%%%%%%%%%%%%%%%%%%%%%%%%%%%%%%%%%%%%%%%%%%%%%%%%%%%%%%%%%%

%%%%%%%%%%%%%%%%%%%%%%%%%%%%%%%%%%%%%%%%%%%%%%%%%%%%%%%%%%%%%%%%%
%    AGRADECIMENTOS
%%%%%%%%%%%%%%%%%%%%%%%%%%%%%%%%%%%%%%%%%%%%%%%%%%%%%%%%%%%%%%%%%

\section{Acknowledgements}

O.R.S. and M.T.T. are in debt with H. Rothe for valuable discussions
about lattice fermionic free models.
O.R.S. thanks CAPES for total financial support.  S.M.de S.  and  M.T.T. 
thank CNPq for  partial  financial support. M.T.T. also
thanks FINEP and  S.M.de S. thanks FAPEMIG  for partial financial 
support.

%%%%%%%%%%%%%%%%%%%%%%%%%%%%%%%%%%%%%%%%%%%%%%%%%%%%%%%%%%%%%%
%       APENDICES
%%%%%%%%%%%%%%%%%%%%%%%%%%%%%%%%%%%%%%%%%%%%%%%%%%%%%%%%%%%%%%

\appendix

\section*{Appendix}

%%%%%%%%%%%%%%%%%%%%%%%%%%%%%%%%%%%%%%%%%%%%%%%%%%%%%%%%%%%%%%
% Formulas uteis
%%%%%%%%%%%%%%%%%%%%%%%%%%%%%%%%%%%%%%%%%%%%%%%%%%%%%%%%%%%%%%

\section{Useful Formulae}

In the  lagrangean density for the free Dirac fermion 
(see eq.(\ref{11})), 

\begin{equation}
{\cal L}=\bar\Psi(\vec x,t)(\partial \hspace{-0.25cm}/ 
 \hspace{0.17cm}  - m)\Psi(\vec x,t),
\label{A1}
\end{equation}

\noindent the Dirac matrices $\gamma^\mu$, $\mu=0,..,3$,  are:

\begin{equation}
\gamma^0 = i   \pmatrix{
                 \1barra  &  \0barra  \cr
                 \0barra   &  -\1barra}
\hspace{1cm}  {\rm and} \hspace{1cm}
\gamma^i = i \pmatrix{ 
               \0barra  & \sigma_i \cr
                - \sigma_i & \0barra},
\hspace{0.5cm} i= 1,2,3.
\label{A2}
\end{equation}

\noindent The matrices $\sigma_i$, $i=1,2,3$ are the Pauli matrices, and
$\1barra$ and $\0barra$ are the identity and null matrices 
of  dimension $2\times 2$ respectively. We have 
 $\bar\Psi(\vec x,t)=-i\Psi_{\nu}^{\dag}(\vec x,t)\gamma_0$.

%\label{A3}

 The spinors components  $u_r(\vec k)$ and  $v_r(\vec k)$
 in eq.(\ref{15}) are:

\vspace{-0.5cm}

\begin{eqnarray}
u_1(\vec k)& =& \sqrt{\frac{m+k_0}{2k_0}}\left(\begin{array}{c}
1\\0\\ \frac{k_3}{m+k_0}\\ \frac{k_1+ik_2}{m+k_0}
\end{array} \right),
\hspace{1cm}
 u_2(\vec k)=
\sqrt{\frac{m+k_0}{2k_0}}\left(\begin{array}{c}
0\\1\\ \frac{k_1-ik_2}{m+k_0}\\ \frac{-k_3}{m+k_0}
\end{array} \right),  \label{A4}  \\
%
%segunda parte da expressao
%
&   &   \nonumber\\
&   &   \nonumber\\
%
%terceira parte da expressao
%
v_1(\vec k)& = &\sqrt{\frac{m+k_0}{2k_0}}\left(\begin{array}{c}
\frac{k_1-ik_2}{m+k_0}\\ \frac{-k_3}{m+k_0}\\
0\\
1
\end{array} \right),
\hskip1cm v_2(\vec k)=\sqrt{\frac{m+k_0}{2k_0}}
\left(\begin{array}{c}\frac{k_3}{m+k_0}\\ \frac{k_1+ik_2}{m+k_0}\\
1\\
0 
\end{array} \right).
\label{A5}
\end{eqnarray}

%%%%%%%%%%%%%%%%%%%%%%%%%%%%%%%%%%%%%%%%%%%%%%%%%%%%%%%%%%%%%%
% Termino das Formulas uteis
%%%%%%%%%%%%%%%%%%%%%%%%%%%%%%%%%%%%%%%%%%%%%%%%%%%%%%%%%%%%%%

\vspace{0.5cm}

%%%%%%%%%%%%%%%%%%%%%%%%%%%%%%%%%%%%%%%%%%%%%%%%%%%%%%%%%%%%%%
% Resultados na rede
%%%%%%%%%%%%%%%%%%%%%%%%%%%%%%%%%%%%%%%%%%%%%%%%%%%%%%%%%%%%%%

\section{Summary of Expressions on the Lattice}

We assume that each direction in the three dimensional space has 
$N$ sites. For simplicity $N$ is assumed to be even. The distance between 
nearest points along each direction is $a$. In the space
lattice, each point is written as $\vec{x} = \vec{n}a$, 
where $\vec{n} = (n_1, n_2, n_3)$, 
$n_i = -\frac{N}{2}, -\frac{N}{2} +1, \cdots, \frac{N}{2}$,
and $i= 1, 2, 3$.

We have the following dictionary between the continous
to the discrete representations  of the formulae:

\begin{eqnarray}
\int_V d^3 \; \vec{x} \hspace{2pt} &  \rightarrow & \hspace{2pt}
  a^3  \hspace{-0.5cm}
\sum_{n_1, n_2, n_3 = -\frac{N}{2}}^{\frac{N}{2}} 
         \label{B1} \\
%
%segunda linha
%
\delta(\vec{x}- \vec{x}^\prime) 
     \hspace{2pt} &  \rightarrow & \hspace{2pt}
\frac{1}{a^3} \delta_{\vec{n}, \vec{m}}^{^{(3)}}
  \equiv \frac{1}{a^3} \delta_{n_1, m_1}
                \delta_{n_2, m_2}  \delta_{n_3, m_3} ,
    \label{B2}
\end{eqnarray}

\noindent where $\vec{x} = \vec{n}a$, 
$\vec{x}^\prime = \vec{m}a$ with 
$\vec{n} = (n_1, n_2, n_3)$ and 
$\vec{m} = (m_1, m_2, m_3)$ respectively.

Finally, the first derivative defined on the lattice
is defined as\cite{rothe}

\begin{equation}
\partial_i \Psi_\alpha (\vec{x}, t) 
\hspace{2pt} \rightarrow \hspace{2pt}
\frac{1}{2a} \left[ \Psi_\alpha (\vec{n}a + \hat{\imath}a, t) 
- \Psi_\alpha (\vec{n}a - \hat{\imath}a, t) \right],
    \label{B3}
\end{equation}

\noindent where $\hat{\imath}$ is the unit vector in the direction 
$i$.

%%%%%%%%%%%%%%%%%%%%%%%%%%%%%%%%%%%%%%%%%%%%%%%%%%%%%%%%%%%%%%
% Termino dos Resultados na rede
%%%%%%%%%%%%%%%%%%%%%%%%%%%%%%%%%%%%%%%%%%%%%%%%%%%%%%%%%%%%%%

%%%%%%%%%%%%%%%%%%%%%%%%%%%%%%%%%%%%%%%%%%%%%%%%%%%%%%%%%%%%%%
%       TERMINO DOS APENDICES
%%%%%%%%%%%%%%%%%%%%%%%%%%%%%%%%%%%%%%%%%%%%%%%%%%%%%%%%%%%%%%

%%%%%%%%%%%%%%%%%%%%%%%%%%%%%%%%%%%%%%%%%%%%%%%%%%%%%%%%%%%%%%
%       REFERENCIAS
%%%%%%%%%%%%%%%%%%%%%%%%%%%%%%%%%%%%%%%%%%%%%%%%%%%%%%%%%%%%%%

\end{document}